\documentclass[symmetry,review,accept,pdftex,moreauthors]{mdpi}

\firstpage{1} 
\pubvolume{1}
\issuenum{1}
\articlenumber{0}
\pubyear{2024}
\copyrightyear{2024}
\externaleditor{Academic Editor: Alberto Ruiz Jimeno}
\datereceived{19 June 2024} 
\daterevised{5 July 2024} 
\dateaccepted{8 July 2024} 
\datepublished{ } 
\hreflink{https://doi.org/\vspace{-12pt}} 

\usepackage{xspace}
\usepackage{footmisc}

\newcommand{\pt}{\mbox{$p_T$}\xspace}

\newcommand{\PP}{\mbox{${\rm I\!P}$}\xspace}

\newcommand{\sqrtnn}{\mbox{$\sqrt{s_{_{NN}}}=$}\xspace}

\newcommand{\jpsi}{\mbox{$J/\psi$}\xspace}

\usepackage{comment}

\Title{SU(3) Gauge Symmetry:  An Experimental Review of Diffractive Physics in $e$$+$$p$, $p$$+$$p$, $p$$+$A, and A$+$A Collision Systems}

\TitleCitation{SU(3) Gauge Symmetry:  An Experimental Review of Diffractive Physics in $e$$+$$p$, $p$$+$$p$, $p$$+$A, and A$+$A Collision Systems}

\Author{{Krista L. Smith} 
 \orcidA}

\AuthorNames{Krista L. Smith}

\AuthorCitation{Smith, K.L.}

\address[1]{
{Los} Alamos National Laboratory, Los Alamos, NM 87545, USA; kristas@lanl.gov\\
}

\abstract{{This review focuses} on diffractive physics, which involves the long-range interactions of strong nuclear force at high energies described by SU(3) gauge symmetry. It is expected that diffractive processes account for nearly 40\% of the total cross-section at LHC energies. These processes consist of soft-scale physics where perturbation theory cannot be applied. Although highly successful and often described as a perfect theory, quantum chromodynamics relies heavily on perturbation theory, a model best suited for hard-scale physics. The study of pomerons could help bridge the soft and hard processes and provide a complete description of the theory of the strong interaction across the full momentum spectrum. Here, we will discuss some of the features of diffractive physics, experimental results from {\mbox{SPS}\xspace}, {\mbox{HERA}\xspace}, and the {\mbox{LHC}\xspace}, and where the field could potentially lead. With the recent publication of the odderon discovery in 2021 by the {\mbox{D0}\xspace} and {\mbox{TOTEM}\xspace} collaborations and the new horizon of physics that lies ahead with the upcoming {Electron-Ion Collider} at Brookhaven National Laboratory, interest is seemingly piquing in high energy diffractive physics. }


\begin{document}


\section{Introduction}

Quantum chromodynamics (QCD) is an SU(3) gauge symmetry \cite{Gross:1973ju} currently used to describe strong interaction, one of the four known fundamental forces in the Universe. In group theory, the special unitary (SU) group is an algebraic group with properties that satisfy Lie algebra.  In the standard model of particle physics, each of the fundamental forces (except gravity) can be described under a unitary group. The electromagnetic force is described by U(1) gauge symmetry, weak interaction is described by SU(2) gauge symmetry, and strong interaction is described by SU(3) gauge symmetry. The standard model of particle physics is then described under the symmetry group  SU(3)$_{c}\times$ SU(2)$_{L}\times$ U(1)$_{Y}$ \cite{Georgi:1974yf}. For a complete description of the standard model and its symmetries, see \cite{Gaillard:1998ui}.

\begin{figure}[H]
\includegraphics[width=9.5 cm]{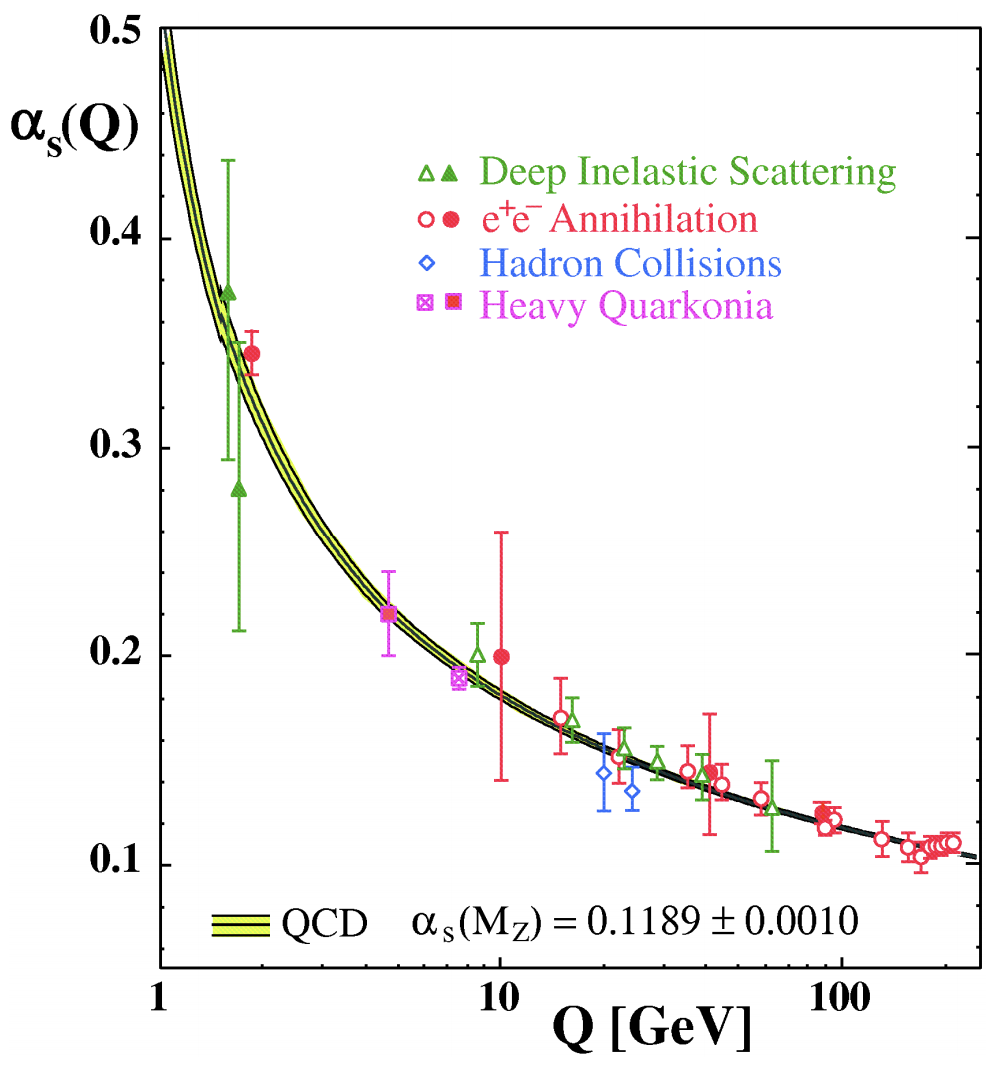}
\caption{\label{fig:alpha_q} {A collection} of experimental measurements of strong coupling $\alpha_{s}(Q)$ are shown as a function of momentum transfer $Q$. Reprinted with permission from Elsevier \cite{Bethke:2006ac}.}
\end{figure}

David Gross and Frank Wilzcek are largely credited for the development of quantum chromodynamics after their joint discovery of asymptotic freedom in 1974 \cite{Gross:1973id}. The idea of asymptotic freedom is well described in Figure \ref{fig:alpha_q}, where the strong coupling $\alpha_{s}(Q)$ is shown as a function of momentum transfer $Q$. The strength of the coupling significantly weakens with increasing momentum transfer. With the reinterpretation of Rolf Hagedorn's maximum temperature for the strong interaction \cite{Hagedorn:1965st}, it was quickly predicted that a new state ``of the vacuum in which quarks are not confined'' \cite{Cabibbo:1975ig} likely exists. Asymptotic freedom then led to wide interest in the discovery of quark-gluon plasma formation in high-energy colliders, which was later confirmed by experiments at the Relativistic Heavy Ion Collider in 2005 \cite{Adcox:2004mh, BRAHMS:2004adc, PHOBOS:2004zne, STAR:2005gfr}. For an overview of QCD in heavy-ion collisions, see \cite{Iancu:1445976}.

\begin{figure}[H]
\includegraphics[width=10 cm]{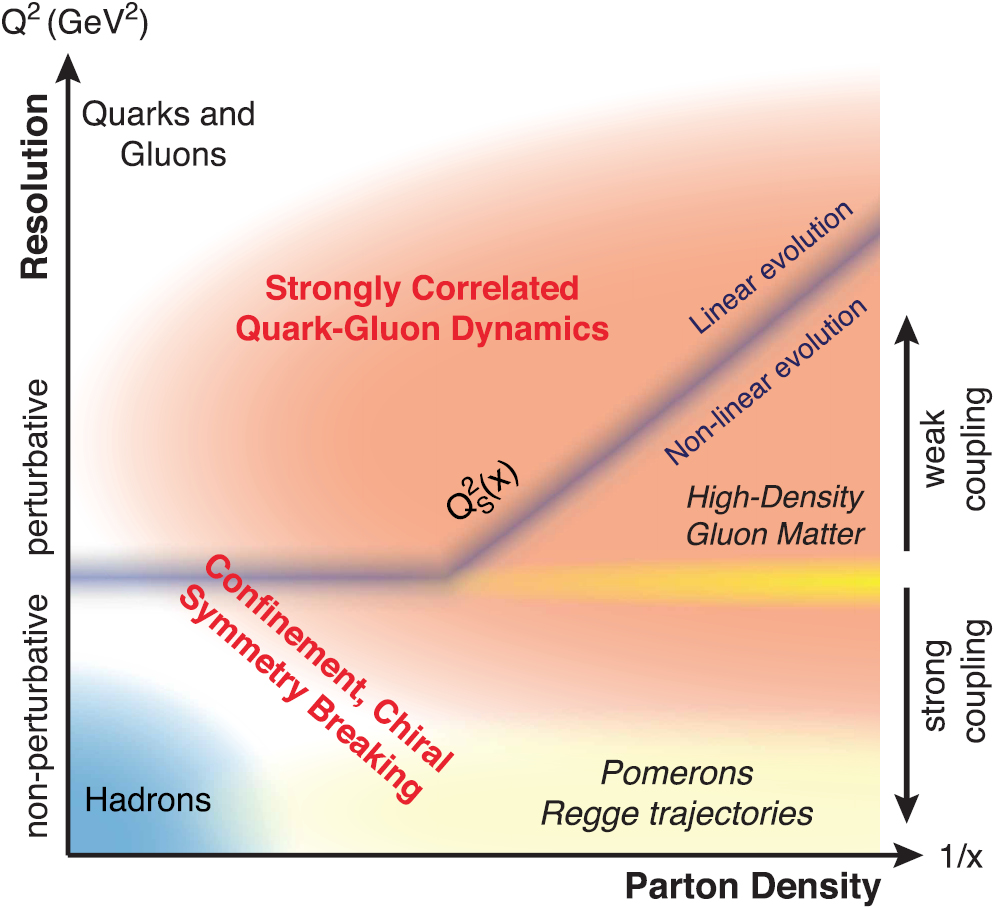}
\caption{\label{fig:eic_pom} Diagram of the squared momentum transfer versus parton density, where pomerons occupy the non-perturbative region of high parton density and strong coupling.  Reprinted with permission from IOP Publishing \cite{Aschenauer:2017jsk}.}
\end{figure}

Although extremely successful and oftentimes described as a perfect theory, it is well known that quantum chromodynamics cannot describe all properties of short and long-range strong interaction. Quantum chromodynamics relies heavily on perturbation theory, which is a model best suited for hard-scale physics. Therefore, the momentum transfer during a collision must be large enough for cross-sections to be calculated using perturbation theory. In this respect, quantum chromodynamics has been hugely successful in describing perturbative physics in the hard-scale regime. However, perturbative processes are estimated to account for $\sim$60\% of the total cross-section \cite{ParticleDataGroup:2020ssz}. The remaining processes consist of soft-scale physics where perturbation theory cannot be applied because the momentum transfer is too small; thus, the strong coupling $\alpha_s$ is too large. Perhaps the study of pomerons, as shown in Figure \ref{fig:eic_pom}, could help bridge the soft (small momentum transfer) and hard (large momentum transfer) processes and potentially provide a complete description of the theory of the strong interaction across the full momentum spectrum.  

Throughout the global development of mathematics and physics over the past few centuries, it seems there is a recurring theme in both areas: the difficulty of working with continuous objects and the desire to discretize these objects. One could think quantum chromodynamics might also incorporate this same recurring theme, where the hard scale can be discretized and has had many successes, while the soft scale remains poorly understood, possibly because it cannot be discretized in the same way. These same challenges also seem to have been present with the advent of quantum field theory proposed by Paul Dirac \cite{Dirac:1928hu} and Hermann Weyl \cite{Weyl:1927vd} in the 1920s, as well as calculus by Isaac Newton \cite{Newton:1687eqk} and Gottfried Wilhelm Leibniz \cite{leibniz} in the 1600s, where methods to discretize continuous objects were successfully found.  

 The Regge model, introduced by Tulio Regge in 1959 \cite{Regge:1959mz}, is a theory for soft physics, where strong interaction is modeled as a continuous exchange of Regge trajectories: ``Regge exchanges provide the binding forces between particles which in their turn generate Regge trajectories'' \cite{Collins:1977jy}. The Regge model gained traction during the 1960s, but in 1971, Leonard Susskind published a paper that suggested moving from modeling the strong interaction with a string model to a discrete chain of partons \cite{Nielsen:1971gkv}, which may have marked the beginning of the field moving away from Regge theory.

In addition to sketching out a unifying theory of the strong interaction, which can describe both hard and soft physics, the study of the soft spectrum might also bring about exciting new physics.  Specifically, looking into a gluon-rich environment such as ultra-peripheral collisions may bring the field closer to new discoveries and possibly glueballs.  Many publications have noted the potential for ultra-peripheral collisions to provide new particle discoveries \cite{Esposito:2021ptx}. With the ALICE FoCal upgrade \cite{ALICE:2020mso, Arslandok:2023utm}, anticipated LHCb Herschel upgrade \cite{Akiba:2018neu}, the CMS and ATLAS Zero Degree Calorimeter upgrades \cite{Bashan:2791533,CMS:2022cju,ATLAS:2781150}, and the building of the EIC \cite{AbdulKhalek:2021gbh}, ultra-peripheral collisions have been garnering more attention in the heavy-ion physics community recently.  

In this review, we focus on diffractive physics, which describes the long-range interactions of strong nuclear force at high energy \cite{Landshoff:1989ku}. The description of diffractive physics relates back to the diffraction pattern seen in optical experiments involving the scattering of light \cite{Abramowicz:1991xz}. Diffractive processes include both elastic and inelastic scattering. In elastic scattering, the target and projectile remain intact after the interaction, while inelastic scattering involves either the target or projectile dissociating or both. Diffractive processes can account for up to 40\% of the total cross-section at LHC energies \cite{ParticleDataGroup:2020ssz}.

Here we will discuss some of the features of diffractive physics, experimental results from {\mbox{SPS}\xspace}, {\mbox{HERA}\xspace}, and {\mbox{LHC}\xspace}, and where the field could potentially lead.  For a complete review of diffractive physics, see \cite{Kaidalov:1979jz,Frankfurt:2013ria}. For a complete review of ultra-peripheral collisions, see \cite{Baltz:2007kq}. For a complete review of semi-hard processes or perturbative processes close to the threshold of soft QCD, see \cite{Gribov:1983ivg}. Finally, for a complete review of central exclusive production, please see \cite{Albrow:2010yb}.  

With the recent publication of the potential odderon discovery in 2021 \cite{D0:2020tig} by the {\mbox{D0}\xspace} and {\mbox{TOTEM}\xspace} collaborations and the new horizon of physics that lies ahead with the upcoming {Electron-Ion Collider} at Brookhaven National Laboratory \cite{Accardi:2012qut, Iancu:2003xm, AbdulKhalek:2021gbh}, interest is seemingly piquing in high energy diffractive physics.

\section{Soft versus Hard Scattering}

\begin{figure}[H]
\includegraphics[width=8.5 cm]{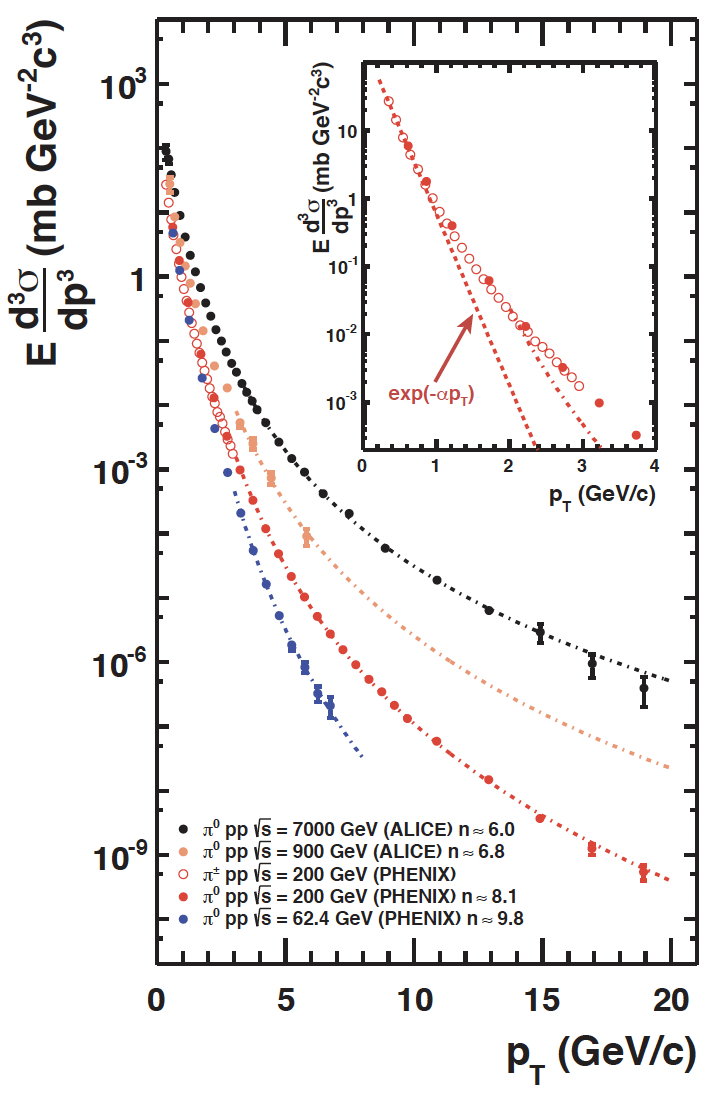}
\caption{\label{fig:regimes} {PHENIX} \cite{PHENIX:2011rvu, PHENIX:2007kqm, PHENIX:2008sgl} and ALICE \cite{ALICE:2012wos} $\pi^{0}$ invariant cross-section measurements in $p$$+$$p$ collisions at different center-of-mass energies are fit with either a power law function ($p_T>2$~GeV/c) or an exponential function ($p_T<1$~GeV/c). The PHENIX experiment recorded data in $p$$+$$p$, $p$$+$A, and A$+$A collisions from 2000$-$2016. Inset: An exponential function cannot describe the data above transverse momentum of $\sim2~$GeV/c. Image Credit: Christian Klein-Boesing, Ph.D. Thesis (University of Munster) \cite{Klein-Boesing:2004fwq}. }
\end{figure}

Soft QCD physics can generally be classified as the energy range in which the transverse momentum of particles produced during a given collision falls below $p_{T}<2~$GeV/c. Equivalently, soft QCD physics describes interactions involving small momentum transfer $Q^{2}<(2~$GeV/c)$^2$. Data collected from the Relativistic Heavy-Ion Collider (RHIC) and the Large Hadron Collider (LHC) shown in Figure \ref{fig:regimes} illustrate the soft and hard QCD physics regimes accessible in today's particle colliders. The $\pi^{0}$ invariant cross-section measured by the PHENIX Experiment at $\sqrt{s}=62.4, 200$~GeV \cite{PHENIX:2011rvu, PHENIX:2007kqm, PHENIX:2008sgl} is compared to measurements by the ALICE Experiment at $\sqrt{s}=900, 7000$~GeV \cite{ALICE:2012wos} as a function of $p_{T}$. The data for $p_{T}>2~$GeV/c are well described by a power law function. However, the cross-section $p_{T}$ dependence changes below this threshold and no longer follows a power law.    

The hard QCD regime has long been favored in high-energy physics, possibly tracing back to the series of hugely successful experiments performed at the Stanford Linear Accelerator (SLAC) in the 1960s. During these years, the SLAC performed deep inelastic scattering experiments that provided compelling evidence the proton is not a point-like object but, in fact, contains a sea of parton constituents \cite{Bjorken:1969ja}. Particles that collide with protons imparting a large enough momentum transfer can, therefore, reveal the internal structure of the proton, the environment where the strong force resides. When exposed to very high energies and very short distances, the strong interaction weakens, and asymptotic freedom prevails \cite{Gross:1973id}, where the bonds of the strong interaction break down, and constituent partons are freed inside a localized region. Hard QCD physics is therefore characterized as ``perturbative'', meaning that perturbation theory \cite{Peskin:1995ev} holds and approximations can be made at short interaction distances based on the weakening of the strong coupling constant $\alpha_s(Q)$ \cite{Bethke:2006ac}.

\begin{figure}[H]
\includegraphics[width=10.0 cm]{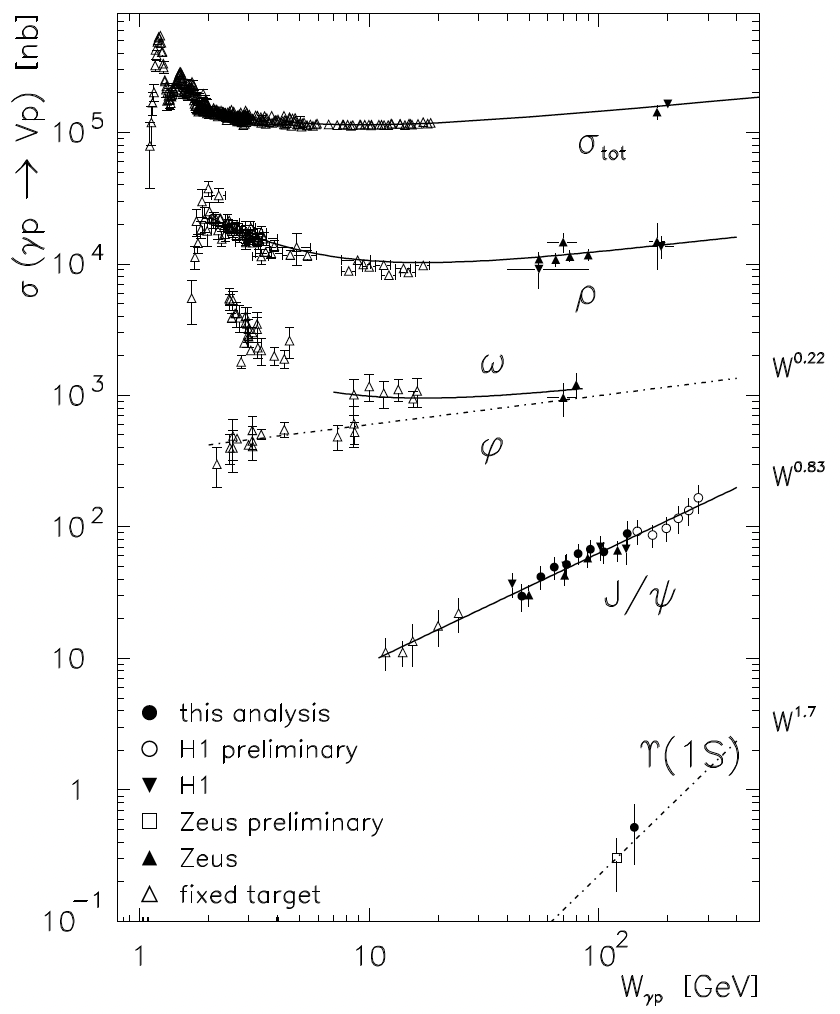}
\caption{\label{fig:merkel_thesis} {From} top to bottom, the cross-section measurements for the $\rho$ meson, $\omega$ meson, $\phi(1020)$ meson, \jpsi meson, and $\Upsilon(1S)$ meson as a function of $W_{\gamma p}$, the total center-of-mass energy of the $\gamma$$-$$p$ system. Image Credit: Petra Merkel, Ph.D. Thesis (University of Hamburg) \cite{Merkel:1999}.}
\end{figure}

Similarly, we see a distinction between a soft and hard scale in the discussion of Regge trajectories and pomeron exchange. Evidence of the pomeron trajectory was determined conclusively \cite{Frankfurt:2013ria} with the H1 \cite{H1:2005dtp} and ZEUS \cite{ZEUS:2002wfj} $e$$+$$p$ collision data from HERA in the early 2000s. Figure \ref{fig:merkel_thesis} presents a striking visual summary of soft versus hard-scale physics using vector meson data measured at HERA. From top to bottom, the cross-sections shown include the following particles, organized from lightest to heaviest: the $\rho$ meson, $\omega$ meson, $\phi(1020)$ meson, \jpsi meson, and $\Upsilon(1S)$ meson. The cross-sections are shown as a function of $W_{\gamma p}$, the total center-of-mass energy of the $\gamma$$-$$p$ system.  We see that the cross-section dependence on $W_{\gamma p}$ becomes more and more pronounced, moving to a larger and larger mass, equivalent to larger $Q^2$ scales and harder scattering, where perturbation theory holds. The distribution shown at the very top of Figure \ref{fig:merkel_thesis} represents the total photo-production cross-section of vector meson measurements. The nonlinear dependence on $W_{\gamma p}$ for the center-of-mass energies roughly below 100 GeV indicates the contribution from soft-scale physics to the total cross-section is indeed sizeable.

\section{Brief Comments on Regge Theory}
As a very brief introduction, Regge theory refers to the use of complex angular momenta as an observable, first postulated by Tulio Regge in `\textit{Introduction to complex orbital momenta}' in 1959 \cite{Regge:1959mz}. This past year (2023) saw more citations of this paper than any other year since its publication, possibly in response to the odderon discovery by the {\mbox{D0}\xspace} and {\mbox{TOTEM}\xspace} collaborations \cite{D0:2020tig}. For a complete review of Regge theory, see \cite{Donnachie:1992ny, 
 Kharzeev:1999vh}. In general, Regge theory holds for small Bjorken-$x$ values \cite{Bjorken:1969ja} (potentially $x < 0.07)$ and for interactions where $W^2$, the total squared center-of-mass energy of the photon-hadron system, is much larger than all other observables \cite{Donnachie:1998gm}.

Here, we will highlight a few of the interesting properties of Regge theory. A Regge trajectory shares some similarities with the idea of a string. In fact, string theory was developed in 1968 by Gabriele Veneziano to explain the strong nuclear interaction and is fundamentally based on Regge theory \cite{Veneziano:1968yb}. However, one significant difference between Regge theory and string theory is that strings can rotate while trajectories do not \cite{Meyer:2004jc}. For a complete review of string theory, see \cite{Schwarz:1998ny}.

\begin{figure}[H]
\includegraphics[width=12 cm]{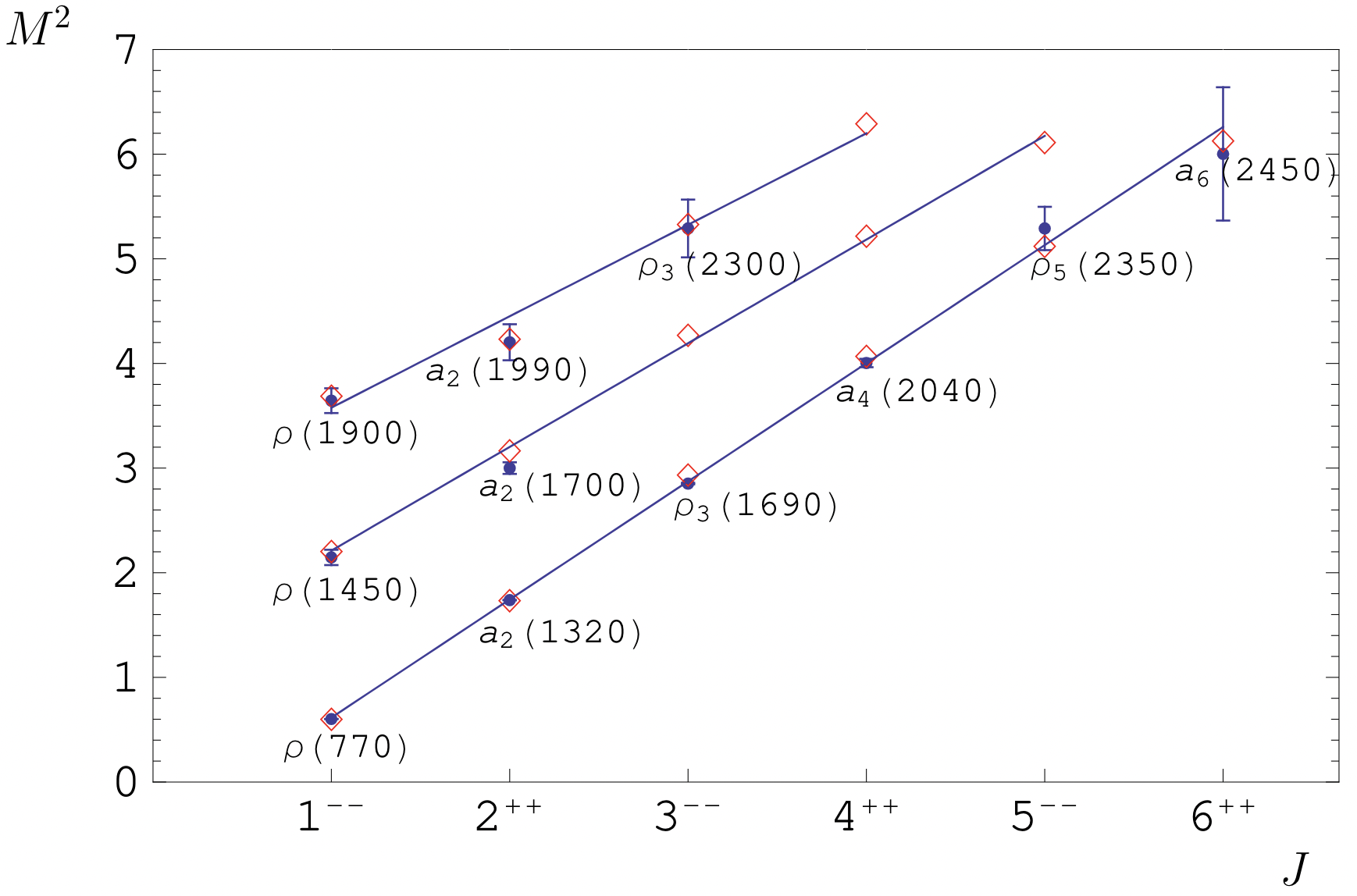}
\caption{\label{fig:regge_line} {The} squared mass $M^2$ versus total spin $J=L+S$ is shown for three families of light unflavored mesons, which all fall along a Regge trajectory. Reprinted with permission from the American Physical Society \cite{Ebert:2009ub}.} 
\end{figure}

Another property of Regge theory is the linear relationship between energy and angular momentum for a $q\bar{q}$ pair:  $E = p + \sigma r$, where $E$ is the total energy, $p$ is the difference in total momentum between the two quarks, $\sigma$ is the string tension, and $r = \sqrt{J/\sigma}$ \cite{Tong2018}.  
A plot of the squared mass versus total spin $J$ \cite{Ebert:2009ub} for $J=L+S$ is shown in Figure \ref{fig:regge_line}, where three Regge trajectories are drawn. Each Regge trajectory begins with a vector meson at the $J^{PC}=1^{--}$ position.  Vector mesons (such as $\phi(1020)$, $\omega$, $\rho$, $\Upsilon$, $J/\psi$, $\psi(2S)$, $\Upsilon(2S)$, etc.) are particles with the same $J^{PC}$ quantum numbers as the photon. Mesons that fall along a Regge trajectory are called a family of mesons. Regge trajectories are also commonly expressed in the form $\alpha(t) = \alpha_{0} + \alpha^{'}(t)$, where $t$ is the square of the momentum transfer, $\alpha_{0}$ is the $y$-intercept, and $\alpha^{'}$ is the slope \cite{Merkel:1999}.

Reggeons are the particles that lie along a Regge trajectory which have a $y$-intercept of less than one \cite{Merkel:1999}. A trajectory with an intercept above one is called the pomeron (\PP) trajectory, first introduced in 1961 by Geoffrey Chu and Steven Frautschi, where the strong interaction involves the exchange of a colorless object known as the pomeron \cite{Chew:1961ev}. The pomeron was proposed by Isaac Pomeranchuk \cite{Berestetsky:1961zat,Gribov:1962gb,Fadin:1975cb,Kuraev:1977fs} and determined conclusively  \cite{Frankfurt:2013ria} with the H1 \cite{H1:2005dtp} and ZEUS \cite{ZEUS:2002wfj} data from {\mbox{HERA}\xspace} in the early 2000s. It has been generally concluded that pomeron exchange occurs in diffractive processes with no quantum number exchange between the target and projectile \cite{Hu:2008zze}, as pomerons are expected to have vacuum quantum numbers (I = S = B = 0; P = C = G = $+$) \cite{Merkel:1999}. Pomerons could be synonymous with the multi-gluon exchange in hadron-hadron scattering \cite{Hu:2008zze} and are often depicted in Feynman diagrams, corresponding to a colorless two-gluon system.  
The pomeron trajectory has long been associated with gluball production \cite{Fiore:2015lnz}, which will be discussed in more detail below. See \cite{Frautschi:1962doa,Kaidalov:1982xe, Kaidalov:1999zb} for a complete discussion of pomerons and soft QCD physics.
 

\section{Experiment Overview}

At the Large Hadron Collider (LHC) in Switzerland, the ALICE \cite{ALICE:2008ngc}, ATLAS \cite{ATLAS:2008xda}, CMS \cite{CMS:2008xjf}, and LHCb \cite{LHCb:2008vvz} experiments have been recording data in $p$$+$$p$, $p$$+$Pb, and Pb$+$Pb collisions since 2010.  
The TOTEM experiment \cite{TOTEM:2008lue} has also been recording data since 2010, but only in $p$$+$$p$ collisions.      
At the Relativistic Heavy-Ion Collider (RHIC) in the United States, the STAR Experiment \cite{STAR:2002eio} has been recording data in $p$$+$$p$, $p$$+$A, and A$+$A collisions since 2000.   
At the Hadron–Electron Ring Accelerator (HERA) in Germany, the H1 \cite{H1:1996prr} and ZEUS \cite{ZEUSCalorimeterGroup:1991wwu} experiments recorded $e$$+$$p$ collision data from {1992$-$2007}. At the Continuous Electron Beam Accelerator Facility (CEBAF) in the United States, the GLUE-X Experiment \cite{GlueX:2020idb} has been recording data since 2014 using a photon beam incident on a fixed target of liquid hydrogen.    
At the Super Proton Synchrotron (SPS) in Switzerland, the WA102 Experiment \cite{Peigneux:1994gy} recorded $p$$+$$p$ collision data from {1995$-$1996}. Lastly, the COMPASS Experiment \cite{COMPASS:2007rjf}, also known as NA58, recorded data from 2002 to 2021 at the SPS.

\section{Potential Sources of Background}

\begin{figure}[H]
\includegraphics[width=13.5 cm]{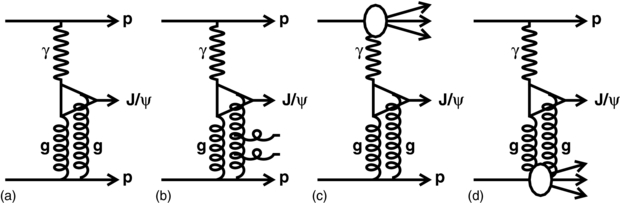}
\caption{\label{fig:cep_feynman} {Central} exclusive production of vector mesons through interactions between a virtual photon and two gluons (or a pomeron \PP). From left to right: (a) elastic interaction, (b) inelastic interaction with gluon radiation, (c) inelastic interaction with target dissociation, and (d) inelastic interaction with projectile dissociation. Reprinted with permission from IOP Publishing \cite{LHCb:2013nqs}.}
\end{figure}

Figure \ref{fig:cep_feynman} illustrates potential sources of background that can be present in, for example, central exclusive production. Vector meson production in $p$$+$$p$ collisions can be understood as the interaction of a virtual photon $\gamma^{\star}$ and a pomeron \PP (denoted by two gluons) between two forward scattered protons. From left to right, the processes shown include elastic interaction with no additional particles produced (a), inelastic interaction with gluon radiation (b), and the target (c) or projectile proton (d) dissociated by the photon-pomeron interaction. The interactions which produce no additional activity (a) are considered signals, while the remaining three processes, (b), (c), and (d), where additional activity is produced during the interaction, are considered backgrounds. 

\begin{figure}[H]
\includegraphics[width=6.5 cm]{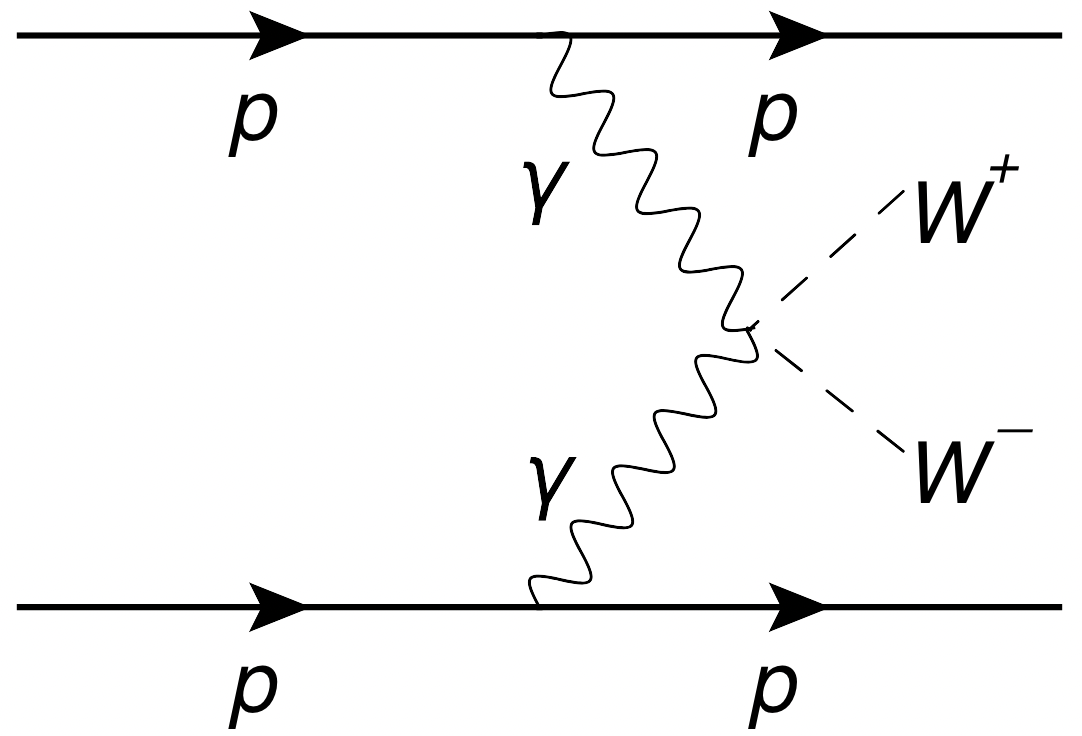}
\caption{\label{fig:wboson} ATLAS experiment observes photon-induced $W^+W^{-}$ production in $p$$+$$p$ collisions at $\sqrt{s}=$~13~TeV~\cite{ATLAS:2020iwi}. Reprinted with permission from Springer \cite{Baldenegro:2020qut}.}
\end{figure}

\begin{figure}[H]
\includegraphics[width=5 cm]{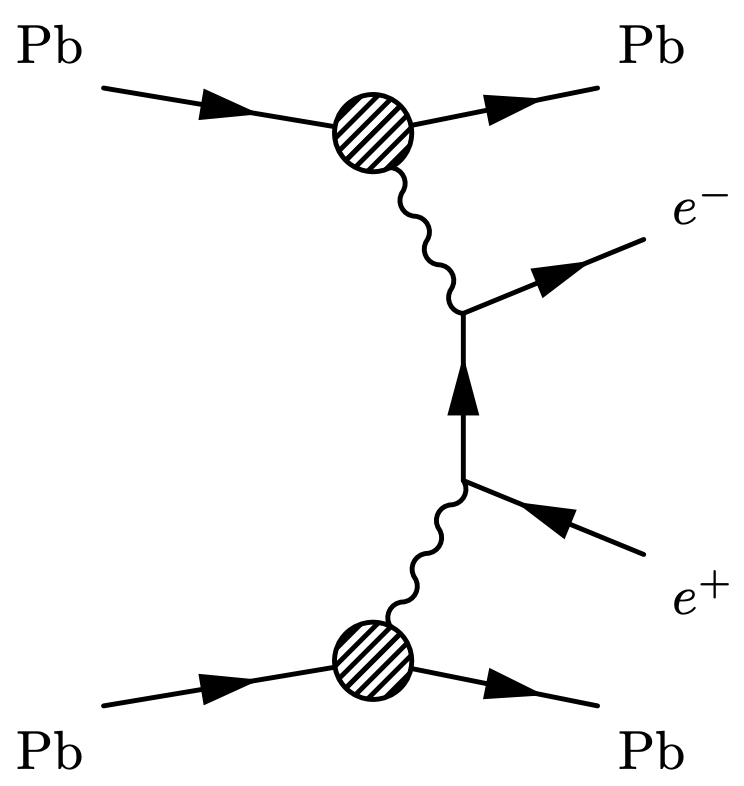}
\caption{\label{fig:ee_feynman} ATLAS experiment observes dielectron production in ultraperipheral
Pb$+$Pb collisions at \sqrtnn 5.02 TeV.  Reprinted with permission from Springer \cite{ATLAS:2022srr}.}
\end{figure}

\begin{figure}[H]
\includegraphics[width=6.5 cm]{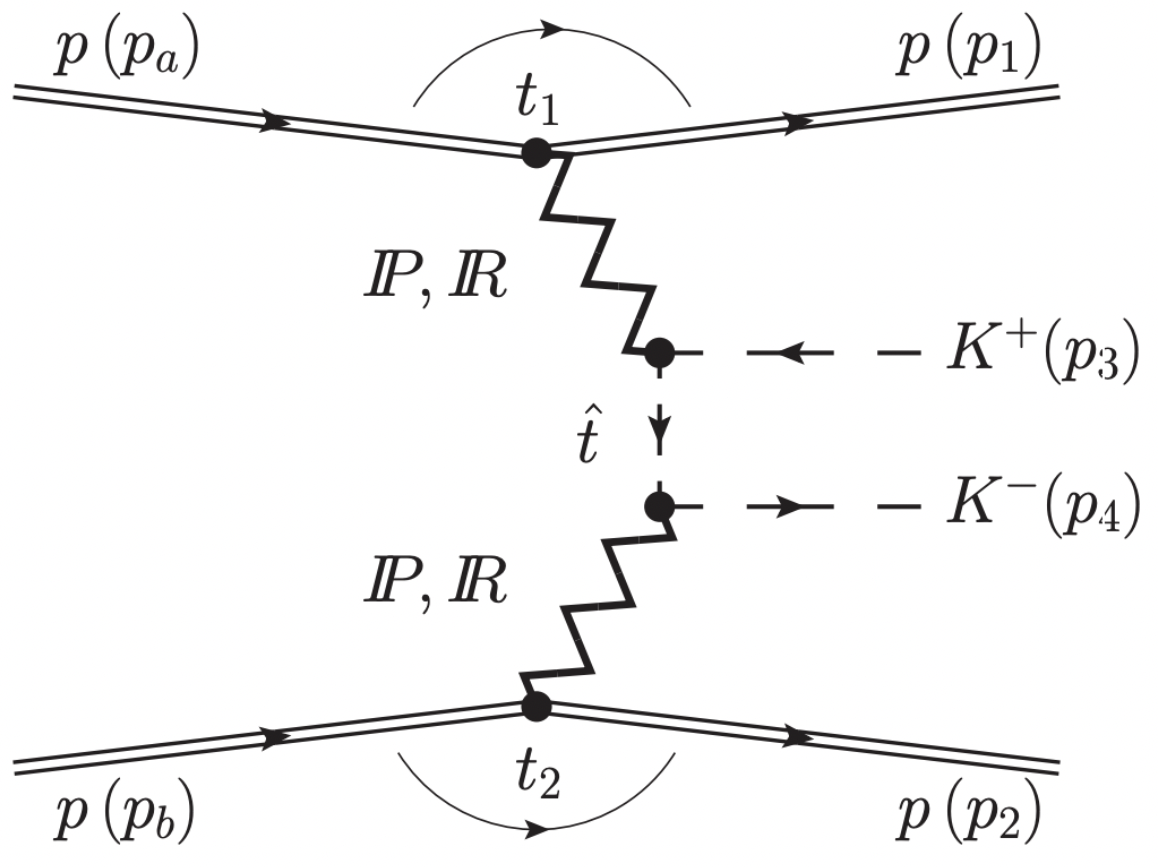}
\caption{\label{fig:kk_feynman} The $\PP$$-$$\PP$ interaction shown is considered non-resonant background in the $K^+K^{-}$ invariant mass distribution. Reprinted with permission from the American Physical Society \cite{Lebiedowicz:2018eui}.} 
\end{figure}

An additional process that can occur both in central exclusive production and ultra-peripheral collisions is the $\gamma\gamma$ interaction. Figure \ref{fig:wboson} shows an ATLAS measurement of photon-induced $W$ boson pair production in central exclusive events at $\sqrt{s}=$ 13 TeV \cite{ATLAS:2020iwi}. Lepton pairs can also be created through $\gamma\gamma$ interactions. 
 Figure \ref{fig:ee_feynman} shows another measurement by the ATLAS experiment of photon-induced $e^+e^{-}$ pairs in ultra-peripheral collisions at $\sqrt{s_{_{NN}}}=$ 5.02 TeV \cite{ATLAS:2022srr}. Like $\gamma\gamma$ interactions, $\PP$$-$$\PP$ interactions are also believed to generate pairs of particles, such as the dikaon pairs in central exclusive events shown in Figure \ref{fig:kk_feynman}. At the time of writing, diakon pair production has only been published in ultra-peripheral collisions by the ALICE collaboration \cite{ALICE:2023kgv}, although predictions have been made regarding the shape of the distribution and potential interference with other resonances \cite{Lebiedowicz:2018eui}. The $\PP$$-$$\PP$ interaction shown in Figure \ref{fig:kk_feynman} is considered a non-resonant contribution in the $K^+K^{-}$ distribution and, therefore, a potential background source.

\vspace{-6pt}
\vspace{-6pt}


\section{Single Pomeron Exchange}
Particle production in ultra-peripheral collisions is generally expected to proceed through $\gamma\gamma$, $\gamma$$-$$\PP$, or $\PP$$-$$\PP$ interactions. In $\gamma$$-$$\PP$ interactions, a pomeron from the target is expected to interact with a virtual photon from the projectile or vice versa. This type of interaction is shown in Figure \ref{fig:ee_feynman}, where vector mesons are generally produced. As previously mentioned, vector mesons have the same quantum numbers as the photon; therefore, this type of interaction is allowed. Particle production of vector mesons then proceeds through single pomeron exchange.

\begin{figure}[H]
\includegraphics[width=12 cm]{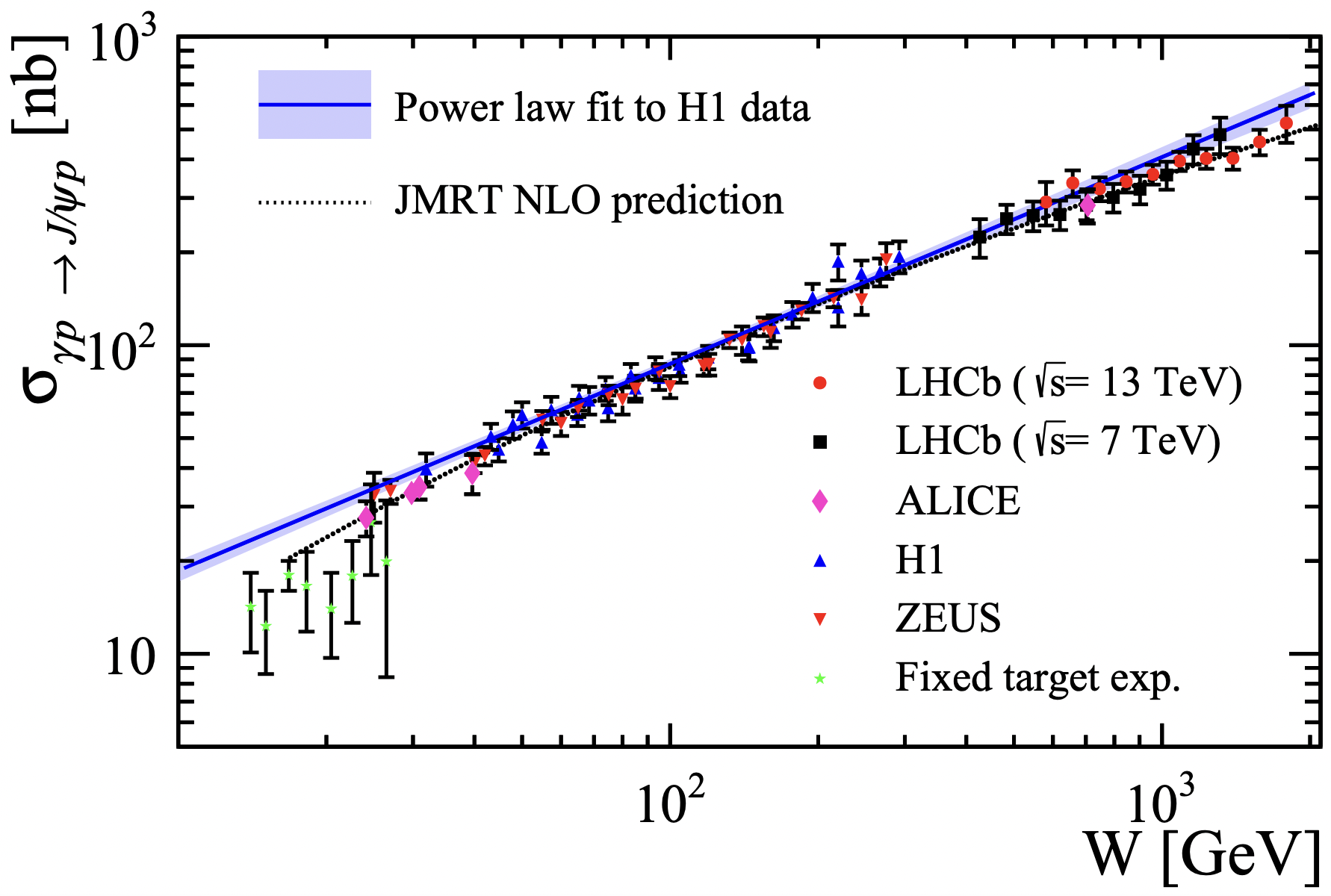}
\caption{\label{fig:lhcb_cep} The LHCb collaboration measured the cross-section for central exclusive production of vector mesons in $p$$+$$p$ collisions as a function of the center-of-mass energy $W_{\gamma p}$ of the $\gamma$$-$$p$ system. The LHCb data are shown for both $\sqrt{s}=7$ TeV (black data points) and $\sqrt{s}=13$ TeV (red data points). Reprinted with permission from Springer \cite{LHCb:2018rcm}. }
\end{figure}

Figure \ref{fig:lhcb_cep} from the LHCb collaboration shows the cross-section for the central exclusive production of vector mesons $J/\psi$ and $\psi(2S)$ in $p$$+$$p$ collisions as a function of the center-of-mass energy $W_{\gamma p}$ of the $\gamma$$-$$p$ system. The LHCb data are measured in both $\sqrt{s}=7$ TeV (black data points) and $\sqrt{s}=13$ TeV (red data points), and results are compared with collider measurements from the ALICE, H1, and ZEUS collaborations as well as fixed target data. The collider data, recorded at center-of-mass energies $W_{\gamma p}$ above 100 GeV, all follow a similar power law trend of increasing cross-section measurements with the increasing center-of-mass energy, as expected with single pomeron exchange. The authors note a potential deviation from the main trendline for the higher $\sqrt{s}=13$ TeV data points.

\begin{figure}[H]
\includegraphics[width=11 cm]{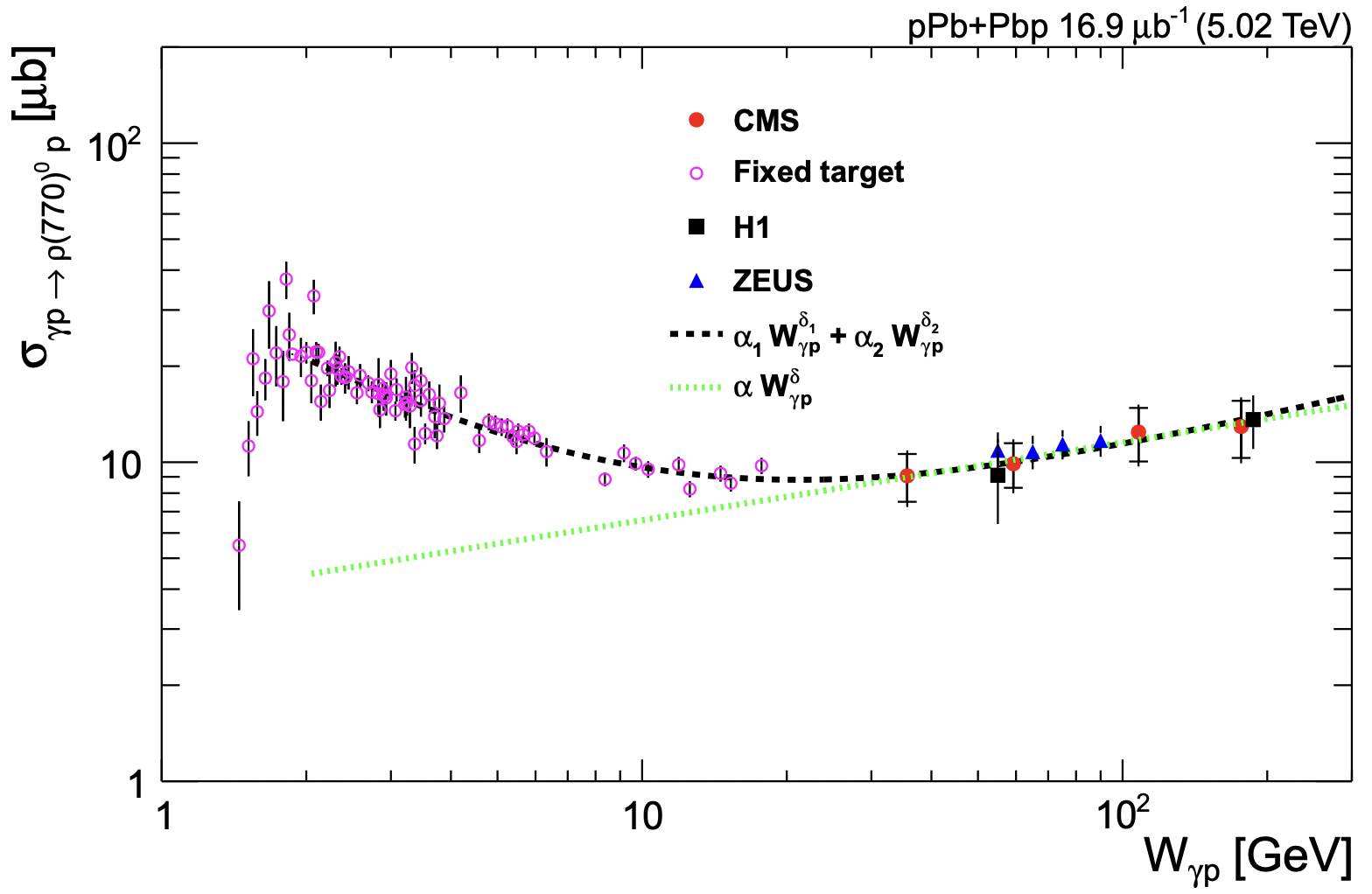}
\caption{\label{fig:cms_rho}{CMS} collaboration results (red solid data points) for the $\rho(770)$ cross-section in $p$$+$Pb collisions as a function of the center-of-mass energy $W_{\gamma p}$ of the $\gamma$$-$$p$ system. The CMS data are compared with fixed target data (magenta open data points), where a clear difference in $W_{\gamma p}$ dependence can be observed. Reprinted with permission from Springer \cite{CMS:2019awk}. \vspace{4mm}}
\end{figure}

In Figure \ref{fig:cms_rho}, the CMS collaboration measures the photo-production cross-section of the vector meson $\rho(770)$ in $p$$+$Pb collisions at $\sqrt{s}=5$ TeV as a function of the center-of-mass energy $W_{\gamma p}$ of the $\gamma$$-$$p$ system. The CMS data are compared with fixed target data, where a clear difference in $W_{\gamma p}$ dependence can be observed. At higher center-of-mass energies, the $\rho(770)$ cross-section approaches the behavior expected from hard scattering. At the lower fixed target energies, a clear distinction can be seen between hard and soft scale physics, where only the hard scale QCD measurements are well described by a power law function. Again, the collider data, taken at $W_{\gamma p}$ energies above 100 GeV, all follow a similar trend.

\begin{figure}[H]
\includegraphics[width=13.5 cm]{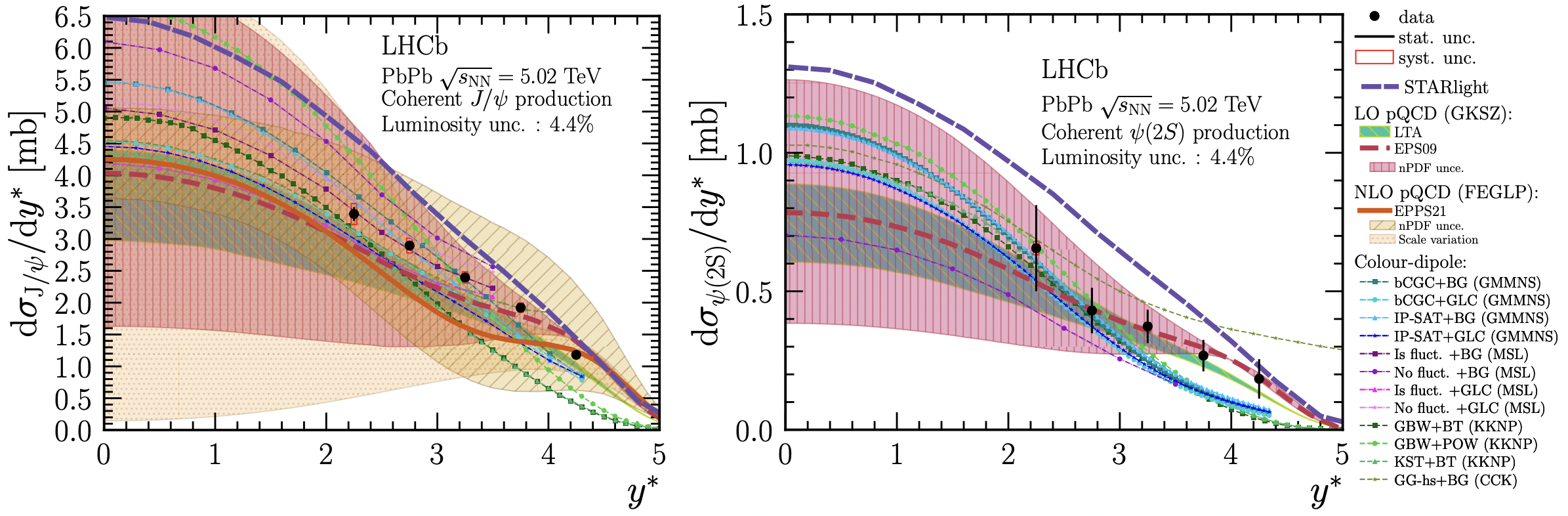}
\caption{\label{fig:lhcb_upc} \hl{The} LHCb collaboration differential cross-section results for $J/\psi$ photo-production (left) and $\psi(2S)$ photo-production (right) as a function of rapidity $y$ in ultra-peripheral Pb$+$Pb collisions at $\sqrt{s_{_{NN}}}=5$ TeV. The data are compared with theoretical predictions based on perturbative quantum chromodynamic  \hl{(see} \cite{Gribov:1983ivg}) and color glass condensate \hl{(see} \cite{{Iancu:2003xm}}) calculations.  Reprinted with permission from Springer \cite{LHCb:2022ahs}.}
\end{figure}

Figure \ref{fig:lhcb_upc} from the LHCb collaboration shows the differential cross-section for $J/\psi$ photo-production (left) and $\psi(2S)$ photo-production (right) as a function of rapidity $y$ in ultra-peripheral Pb$+$Pb collisions at $\sqrt{s_{_{NN}}}=5$ TeV. The differential measurements for $J/\psi$ and $\psi(2S)$ have a similar dependence on rapidity, both decreasing with increasing rapidity.  However, the authors note that the decreasing trend is not constant, and a small bump can be observed between rapidity 3 and 4. The cross-section measurements are compared to theoretical predictions based on color glass condensate (CGC) and perturbative quantum chromodynamics (pQCD) calculations. The data appear to be best described by pQCD calculations.  

 \vspace{-12pt}

\section{Double Pomeron Exchange}
As previously described, the photo-production process during which a single virtual photon and a single pomeron are exchanged between the target and projectile nuclei is known as single pomeron exchange. This process could also be described as a single diffractive process, which implies a single photon or pomeron exchange during the interaction. Similarly, if two virtual photons or two pomerons are exchanged during an interaction, the process could be described as double diffractive. If the process involves the exchange of two pomerons, it would then be considered a double pomeron exchange.

Fewer experimental results are available for double pomeron exchange than single pomeron exchange, likely due to the design of experimental detectors and the lack of far-forward rapidity coverage. In a 2010 review of double pomeron exchange experimental results \cite{Albrow:2010zz}, it was noted that the rapidity gap expected between $X$, the center-of-mass energy of particles produced near central rapidity, and the projectile nucleus is $\Delta y = 3$.  At that time, it seems none of the major LHC experiments had detector coverage capable of measuring this kind of rapidity range. In experiments prior to 2010, the Colliding Detector at Fermilab (CDF), for example, had poor statistics, mass resolution, or no trigger available to select double pomeron exchange events \cite{Albrow:2010zz}. It was also mentioned that some experiments focused more on other physics results, such as jet measurements. A more recent summary of experimental and theoretical studies of double pomeron exchange can be found in \cite{Lebiedowicz:2018eui}.

\begin{figure}[H]
  \includegraphics[width=13.75 cm]{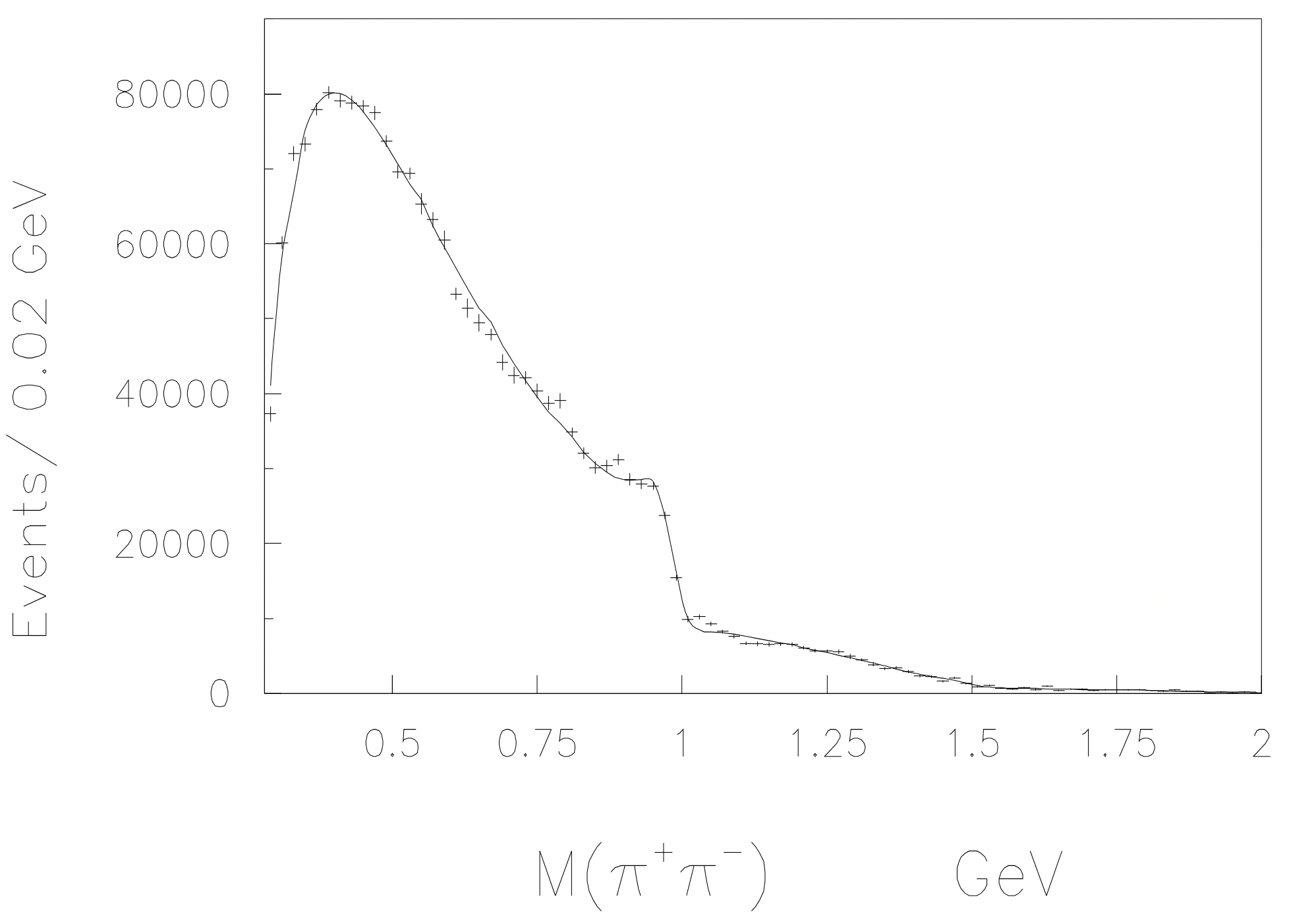}
  \makebox[0pt][r]{
    \raisebox{6.5cm}{
      \includegraphics[width=7.75 cm]{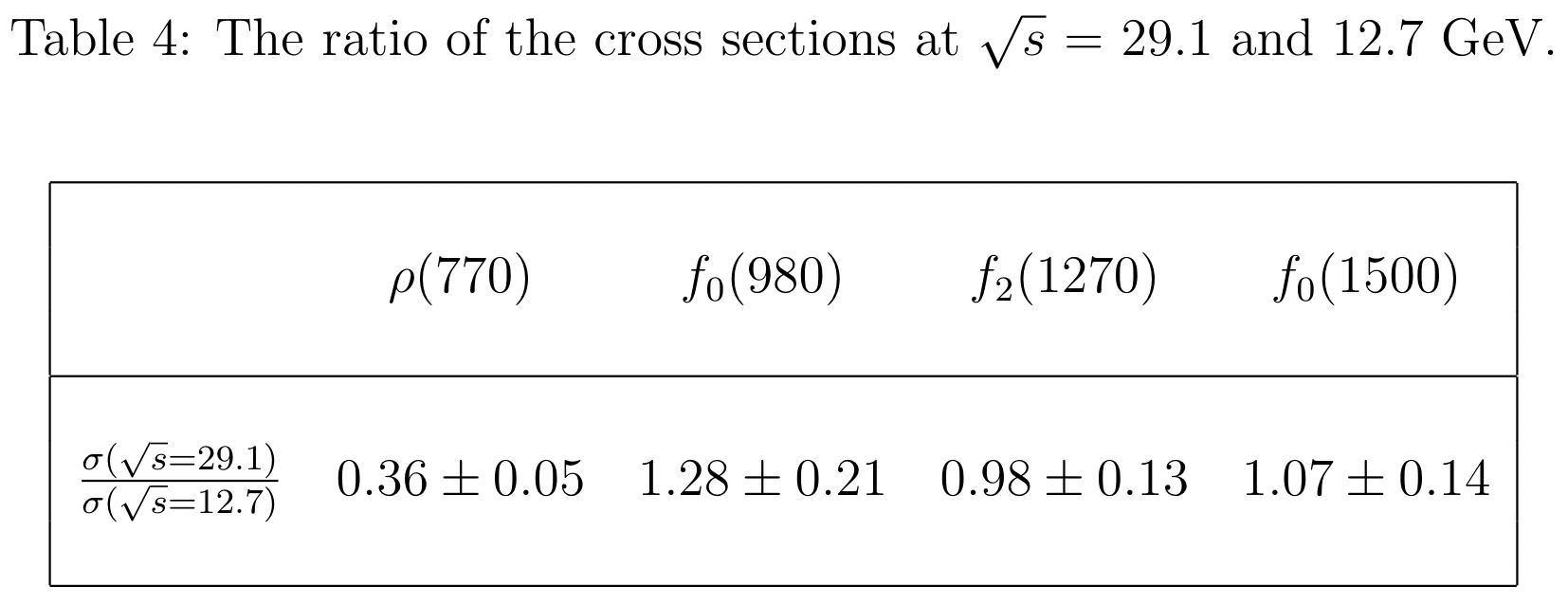}
    }\hspace*{0.5cm}
  }
  \caption{\label{fig:wa102} \hl{The} $\pi^+\pi^{-}$ invariant mass spectrum is shown for central production in $p$$+$$p$ collisions at 450 GeV/$c$ \hl{incident beam momentum} by the WA102 collaboration. Inset: Measured cross-section ratios for the $\rho(770)$, $f_{0}(980)$, $f_{2}(1270)$, and $f_{0}(1500)$ resonances recorded at two different collision energies. Reprinted with permission from Elsevier \cite{WA102:1999fqy}.}
\end{figure}

Figure \ref{fig:wa102} from the WA102 collaboration shows the $\pi^+\pi^{-}$ invariant mass spectrum for central production in $p$$+$$p$ collisions at 450 GeV/$c$. The $\rho(770)$, $f_{0}(980)$, $f_{2}(1270)$, and $f_{0}(1500)$ resonances appear in the mass spectrum, and the fully corrected cross-sections are shown in the inset table for two different collision energies. The resulting ratios consist of unity for the $f_0$ and $f_2$ mesons, unlike the $\rho$ meson, indicating the three $f_0$ and $f_2$ resonances are produced through double pomeron interactions. See \cite{Ganguli:1980mp} for more details on why cross-section measurements that are independent of energy are consistent with double pomeron exchange. 

\begin{figure}[H]
\includegraphics[width=11 cm]{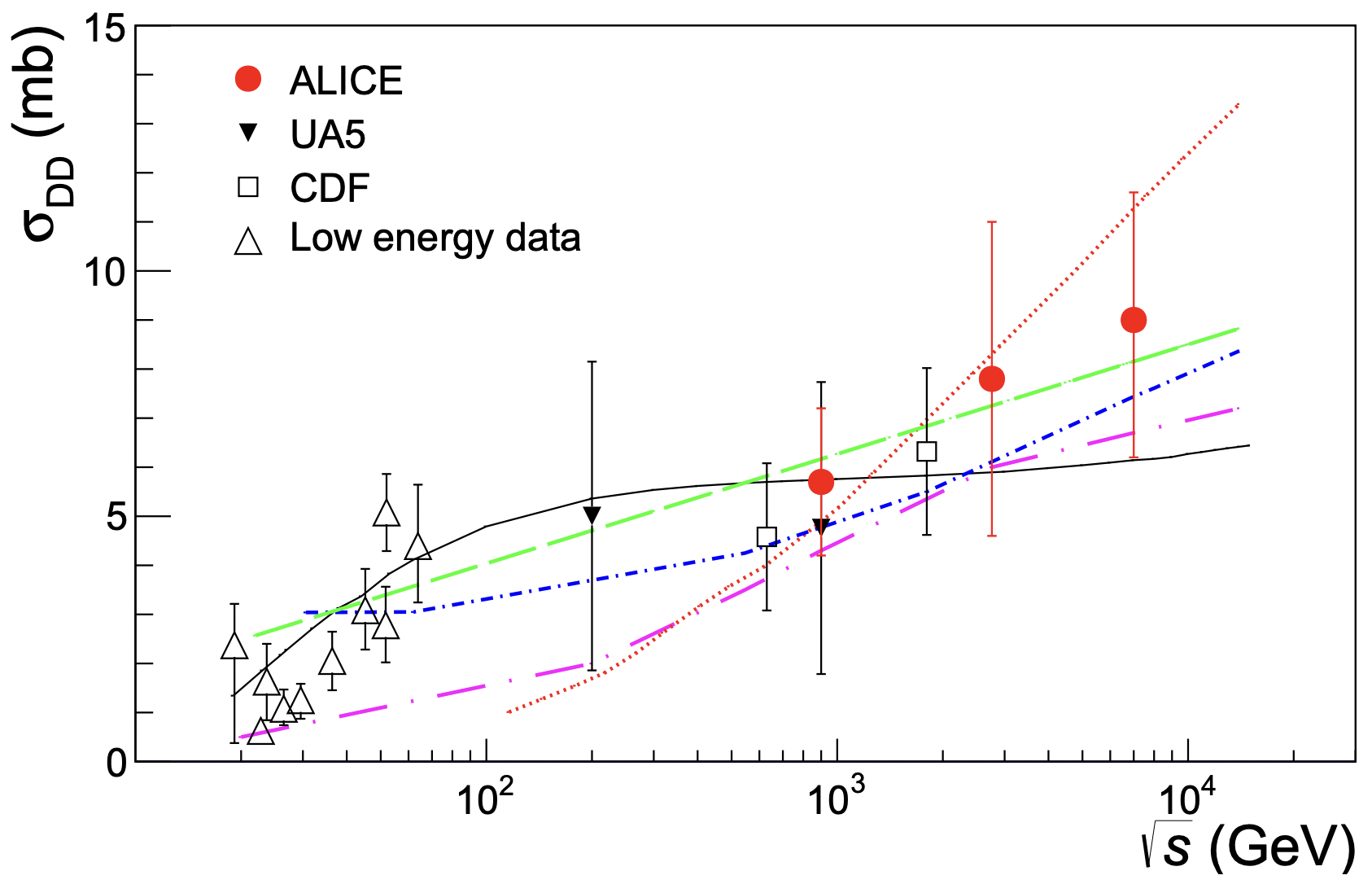}
\caption{\label{fig:alice_DPE}{The} ALICE collaboration results (red data points) for the cross-section of charged particles produced through double pomeron exchange as a function of center-of-mass energy $\sqrt{s}$. The ALICE LHC data are compared with lower energy collider data. Reprinted with permission from Springer \cite{ALICE:2012fjm}.}
\end{figure}

Figure \ref{fig:alice_DPE} from the ALICE collaboration shows the cross-section for charged particle production through double pomeron exchange as a function of center-of-mass energy $\sqrt{s}$. The results from ALICE are compared to other measurements taken at lower collision energies, including the UA5 and CDF collaborations. Although the uncertainties are generally more significant for the higher energy measurements, the distribution as a function of collision energy is relatively flat, especially at energies above $\sqrt{s}=100$ GeV. The authors conclude that measurements below 100 GeV collision energy are also consistent with the higher energy data and, therefore, that the cross-section for particles produced through double pomeron exchange is, in fact, independent of collision energy. 

\begin{figure}[H]
\includegraphics[width=8 cm]{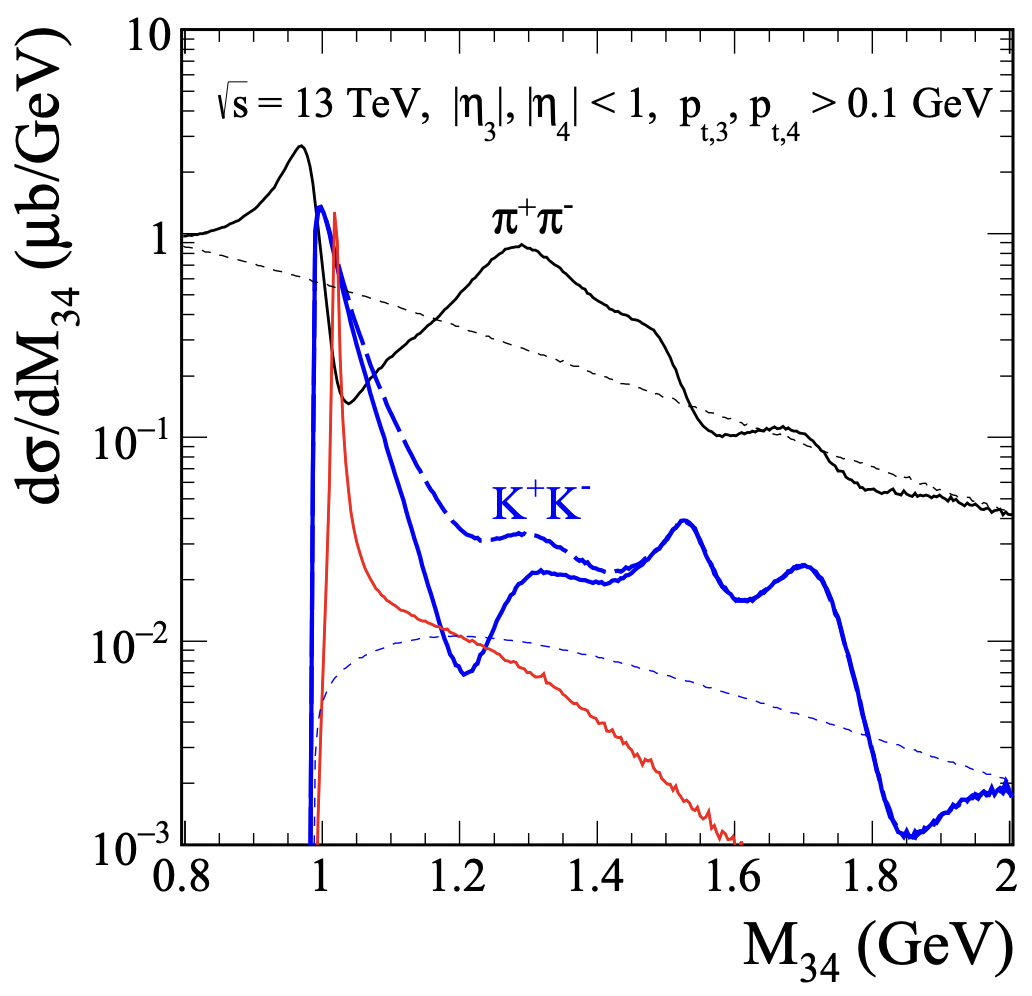}
\caption{\label{fig:kk_nonres} {Predictions} of the $K^+K^{-}$ invariant mass spectrum for central production in $p$$+$$p$ collisions at $\sqrt{s}=13$ TeV. The resonances shown by the blue curve include scalars $f_0(980)$, $f_0(1500)$, $f_0(1710)$, and the tensors $f_2(1270)$ and $f_2^{'}(1525)$, and are produced via double pomeron exchange. The curve shown in red is the expected contribution from photo-production, which includes the vector meson $\phi(1020)$. Reprinted with permission from the American Physical Society \cite{Lebiedowicz:2018eui}.}
\end{figure}

Figure \ref{fig:kk_nonres} shows predictions of the $K^+K^{-}$ invariant mass spectrum for central production in $p$$+$$p$ collisions at 13 TeV, assuming double pomeron exchange. The resonances shown by the blue curve include scalars $f_0(980)$, $f_0(1500)$, $f_0(1710)$, and the tensors $f_2(1270)$ and $f_2^{'}(1525)$, and are produced via double pomeron exchange. The curve shown in red is the expected contribution from photo-production, which includes the vector meson $\phi(1020)$. Therefore, the predictions here expect several unflavored light $f_0$ and $f_2$ mesons produced through double pomeron exchange in the $K^+K^{-}$ invariant mass spectrum and that double pomeron exchange occurs alongside single pomeron exchange. Additionally, a sizeable non-resonant continuum is predicted in the $K^+K^{-}$ spectrum.  

\begin{figure}[H]
\includegraphics[width=8.0 cm]{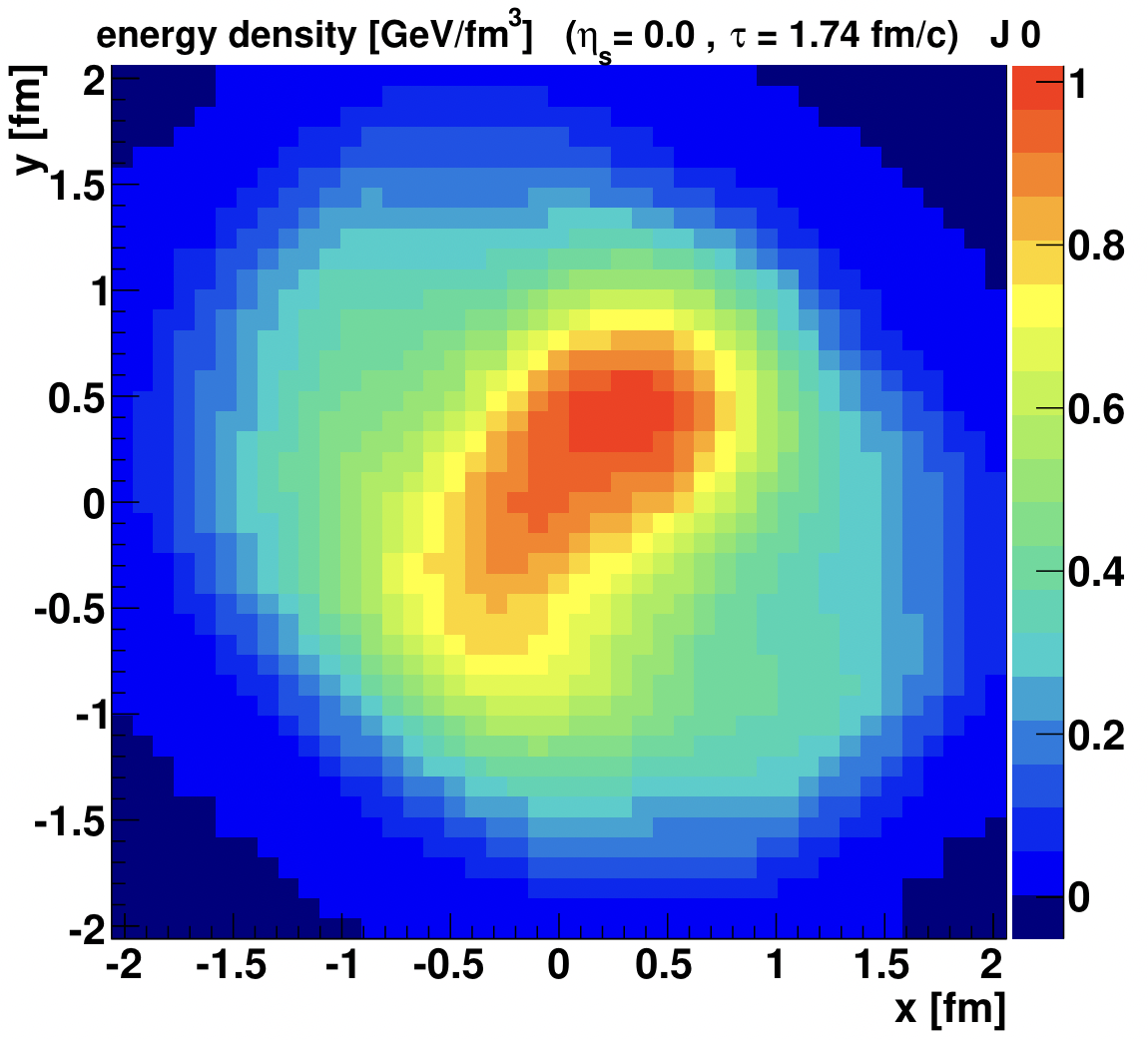}
\caption{\label{fig:epos4_pol} {Predictions} of the energy density formed in $p$$+$$p$ collisions at the LHC in the transverse plane. The model is based on the EPOS4 framework and includes the interactions of six pomerons. Reprinted with permission from the American Physical Society \cite{Zhao:2023ucp}.} 
\end{figure}

Figure \ref{fig:epos4_pol} shows predictions of the energy density in the transverse plane available for particle production in $p$$+$$p$ collisions at LHC energies. This model is based on the EPOS Monte Carlo event generator package, first developed around 2008, which includes $\PP$$-$$\PP$ exchange in terms of parton ladder splitting \cite{Werner:2008zza}. It was perhaps initially believed that double parton exchange would play a more critical role in the physics of larger heavy ion collision systems, such as $p$$+$A or AA collisions, as opposed to $p$$+$$p$ collision systems. However, a recent publication involving the possible formation of quark-gluon plasma in $p$$+$$p$ collision systems \cite{Zhao:2023ucp} with the updated EPOS4 framework \cite{Werner:2023zvo} includes interactions of six pomerons, whereas two pomerons are generally considered in $p$$+$$p$ collisions. The predicted energy density in certain regions of the transverse plane can reach the threshold of $\sim$1~GeV/fm$^{3}$ expected for the formation of quark-gluon plasma \cite{Shuryak:1980tp, Adcox:2004mh}. These results seem to suggest that considering $\PP$$-$$\PP$ interactions in heavy-ion collisions can have measurable effects on the implications of quark-gluon plasma formation and our understanding of asymptotic freedom.

\section{Scalar Glueball Predictions}

Significant efforts have been made towards the understanding and experimental discovery of the scalar glueball, with $J^{PC}=0^{++}$, since at least the 1990s. The WA102 experiment, which collected $p$$+$$p$ data with the SPS from 1995-1996, searched for glueballs in the pseudo-scalar $K^{+}K^{-}$, $\pi^+\pi^{-}$, and $\eta\eta$ decay channels. The X(1750), observed by the WA102 collaboration \cite{WA102:1996vwe}, was initially considered a potential scalar glueball candidate. The STAR experiment also searched for this same signal in ultra-peripheral Au$+$Au collisions \cite{Warnasooriya:2003xza} a few years later but could not find anything, studying both the $K^+K^{-}$ and $\pi^+\pi^{-}$ invariant mass spectra for \pt$<$ 150~MeV/$c$. According to lattice QCD predictions, the scalar glueball should fall within the range of 1-2~GeV/$c^2$ \cite{Vento:2015yja, 
 Sarantsev:2021ein}. An early estimate \cite{Sexton:1995kd} predicts mass $M=1740\pm71$~MeV and width $\Gamma=108\pm28$~MeV. The $f_0$ series of resonances is widely believed to be either pure or hybrid glueballs, and the $f_{0}(1500)$ particle, in particular, has been considered a potential glueball candidate for many years \cite{Amsler:1995td, Crede:2008vw}.  

\begin{figure}[H]
\includegraphics[width=13 cm]{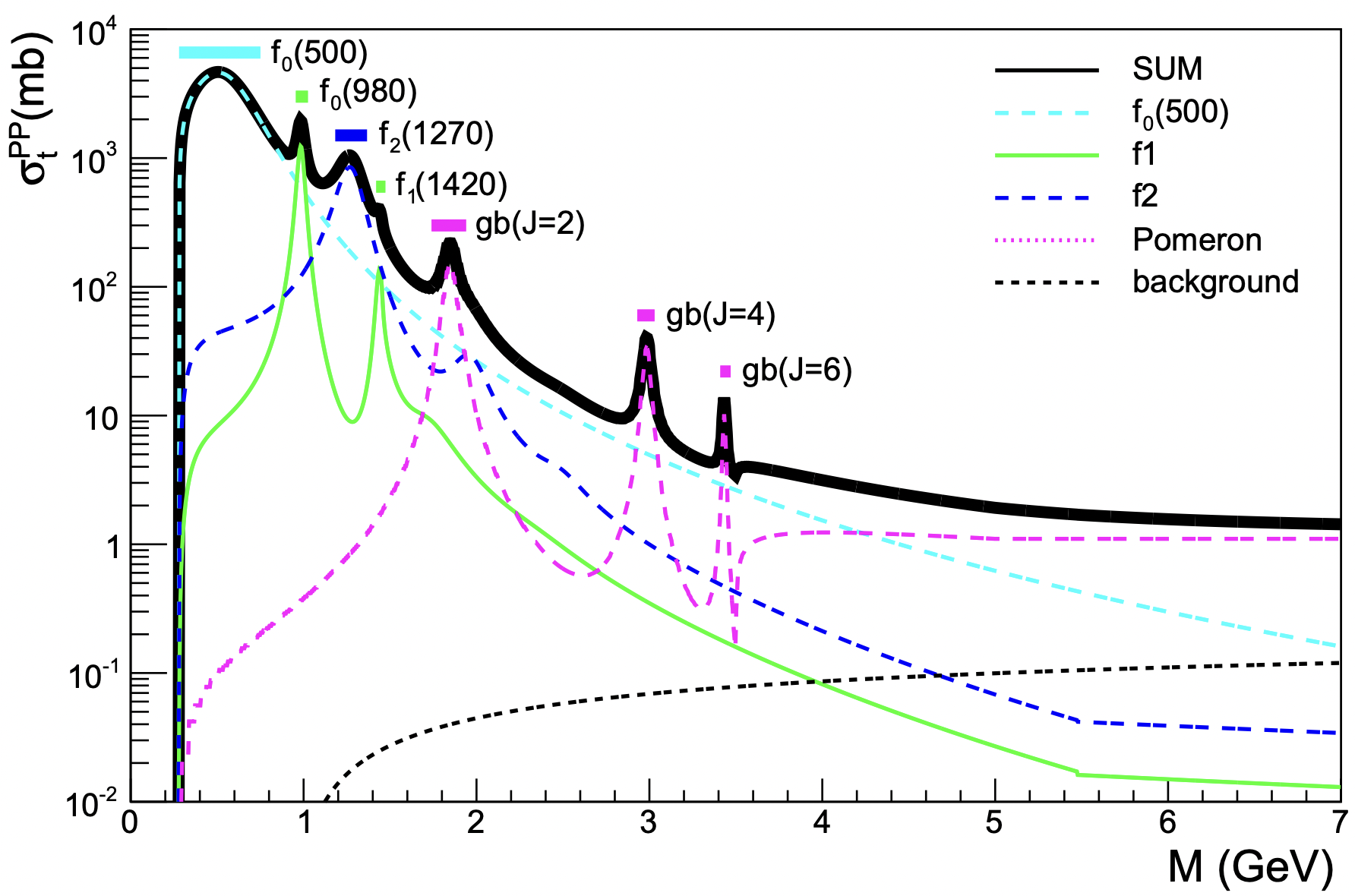}
\caption{\label{fig:kk_glue} {The} predicted Regge trajectories for $f_0(500)$, $f_0(980)$, $f_2(1270)$, and $f_1(1450)$ are shown for the $\pi^+\pi^{-}$ invariant mass spectrum in $p$$+$$p$ collisions at $\sqrt{s}=13$ TeV. The predicted pomeron trajectory (dashed magenta curve) is shown with the corresponding glueball states (gb). Reprinted with permission from Springer \cite{Fiore:2015lnz}.}
\end{figure}

Glueballs could potentially be produced through $\gamma\gamma$ or $\PP$$-$$\PP$ interactions. According to some predictions \cite{Schramm:1999tt}, glueballs produced through $\gamma\gamma$ interactions are expected to have a width on the order of $\sim$ 4 eV, more narrow than the \jpsi width at $\sim$ 93 keV \cite{ParticleDataGroup:2018ovx}. The estimate for a glueball produced through $\PP$$-$$\PP$ interactions, however, is expected to be around 70 MeV or larger. Note these quoted predictions specifically consider the $f_{0}(1710)$ resonance to be a potential scalar glueball candidate. In Figure \ref{fig:kk_glue}, a prediction for the pomeron trajectory (magenta dashed curve) is shown in the $\pi^+\pi^{-}$ invariant mass spectrum. The spectrum includes the $f_{0}(500)$, $f_0(980)$, $f_2(1270)$, and $f_1(1420)$ resonances, where a different trajectory is predicted for each spin $J$. The pomeron trajectory is predicted to begin at the $J=2$ glueball resonance and includes the $J=4$ and $J=6$ glueball states. The scalar glueball is not expected to lie on the pomeron trajectory \cite{Meyer:2004jc}.  

After many years of searches and debates, there remains no firm candidate for the scalar glueball. However, on 2 May 2024, \textit{Physical Review Letters} published a result from the BESIII Collaboration on the $X(2370)$ as a potential pseudoscalar glueball state with $J^{PC}=0^{-+}$ \cite{BESIII:2023wfi}. This article has been featured as an Editors' Suggestion, and as of 5 July 2024, it is within the top 5\% of all research outputs scored by Altmetric \cite{Alt:2024}. Based on the interest surrounding these results, the physics community is eager for a glueball discovery, and the $X(2370)$ currently looks to be a plausible candidate.


\section{Conclusion} 

In summary, experimental results from single pomeron exchange reveal a clear distinction between soft and hard scale physics. The measurements of the $J/\psi$, $\psi(2S)$, and $\rho(770)$ vector mesons as a function of $W_{\gamma p}$, the center-of-mass of the $\gamma$$-$$p$ system, all follow a similar power law trend from above roughly 100 GeV. Additionally, the experimental data collected at varying energies by the different experimental collaborations (H1, ZEUS, ALICE, CMS, and LHCb) are consistent between LHC and HERA energies. However, the measurements of the $\rho(770)$ meson recorded at fixed target energies clearly do not follow this same power law dependence.

From the WA102 (c. 1999) and ALICE (c. 2013) results for double pomeron exchange and the more recent theoretical predictions (c. 2018) for DPE, a consistent picture emerges among them. Light, unflavored scalar ($f_0$) and tensor ($f_2$) mesons are expected to be produced through double pomeron exchange. Additionally, if the production mechanism is indeed double pomeron exchange, the measured cross-section for these resonances should be independent of energy, certainly above 100 GeV collision energy. From the presence of the $\rho(770)$ and $\phi(1020)$ vector mesons alongside the unflavored mesons in the same mass spectrum, there is also a consistent picture that single pomeron exchange (photo-production) occurs alongside and seemingly independently from double pomeron exchange. Additionally, the predictions of energy density present in $p$$+$$p$ collisions at the LHC could be correlated with $\PP$$-$$\PP$ interactions. It appears that if more $\PP$$-$$\PP$ interactions are included in a model than typically expected, the energy density levels can reach the threshold widely believed necessary for quark-gluon plasma formation.     

\section{Future Directions}
With the ALICE FoCal upgrade \cite{ALICE:2020mso, Arslandok:2023utm}, the potential LHCb Herschel upgrade \cite{Akiba:2018neu}, the CMS and ATLAS Zero Degree Calorimeter upgrades \cite{Bashan:2791533,CMS:2022cju,ATLAS:2781150}, and the building of the {Electron-Ion Collider} \cite{AbdulKhalek:2021gbh}, ultra-peripheral collisions and central exclusive production have been garnering more attention in the field of high energy physics and more specifically in the heavy-ion community. As previously mentioned, the {\mbox{D0}\xspace} and {\mbox{TOTEM}\xspace} collaborations in 2021 jointly discovered the odderon \cite{D0:2020tig}, and citations of Tulio Regge's first paper on Regge theory \cite{Regge:1959mz} have since seen more citations in the year 2023 than any other year since its publication more than fifty years ago in 1959.

Interest has seemingly been renewed in diffractive physics, with numerous experimental upgrades projected for the four major experiments at the LHC and the construction of the new {Electron-Ion Collider} at Brookhaven National Laboratory expected to begin in the 2030s. It seems entirely possible that new discoveries might lay ahead for diffractive physics and possibly in the field of heavy-ion physics.  Heavy-ion collisions could potentially provide a gluon-rich environment to study the long-range interactions of double pomeron exchange.  It has also been speculated there could be an entirely separate fundamental force that describes the strong interaction at long-range distances \cite{Landshoff:1989ku}.    

\vspace{6pt}
\funding{Los Alamos National Laboratory is supported by the United States Department of Energy/Office of Science/Office of Nuclear Physics.}

\dataavailability{Most of the data results reviewed in this article from the ALICE, ATLAS, CMS, H1, LHCb, and ZEUS Collaborations are available at the HEPData database of high energy physics experimental scattering data at {\url{https://www.hepdata.net/} (accessed on 25 June 2024). }}

\acknowledgments{All figures in this article have been previously published, and the author wishes to thank all original authors for their work.}

\conflictsofinterest{The authors declare that this study received funding from Knowledge Unlatched, a Wiley brand and company of John Wiley \& Sons, Inc. The funder had the following involvement with the study: Through an Open Access agreement between Knowledge Unlatched and the Los Alamos National Laboratory Research Library, the Article Processing Charge for publication was funded by Knowledge Unlatched in support of Open Access publishing for Los Alamos National Laboratory authors. The funder was not involved in the study, design, collection, analysis, interpretation of data, the writing of this article, or the decision to submit it for publication.} 


\begin{adjustwidth}{-\extralength}{0cm}

\reftitle{References}


\end{adjustwidth}
\end{document}